%% file: slemc.tex
\documentclass[9pt,twocolumn]{IEEEtran} 

\def\BibTeX{{\rm B\kern-.05em{\sc i\kern-.025em b}\kern-.08em
    T\kern-.1667em\lower.7ex\hbox{E}\kern-.125emX}}

\usepackage{times}
\usepackage{graphics}
\usepackage{multicol}
\RequirePackage{float}
\usepackage{here}
\usepackage{algorithm}
\usepackage{algorithmic}
\usepackage{amsmath}
\usepackage{amsfonts}
\usepackage{amsbsy}
\usepackage{amssymb}
\usepackage{rotating}
\usepackage{wrapfig}
\usepackage{subfigure}
\usepackage{mathptm}
\usepackage{mathdots}
\usepackage{fancybox}
\usepackage{epic}
\usepackage{eepic}

\DeclareMathVersion{bold}

\include{boxedeps}
\SetRokickiEPSFSpecial
\HideDisplacementBoxes

\def\ssp{\def\baselinestretch{0.93}\large\normalsize}

\input{header}

\input{symbols}
\input{commands}

\title{  {
\LARGE \bf
Fast Monte Carlo Estimation of Timing Yield:\\
\vspace{-.25cm} Importance Sampling with Stochastic Logical Effort
(ISLE) } }

\author{\Large Alp Arslan Bayrakci $\quad\quad$ Alper Demir $\quad\quad$ Serdar Tasiran\\
\vspace{0.2cm} \thanks{This work was funded by TUBA (Turkish
Academy of Sciences) under the GEBIP Program and by TUBITAK
(Scientific and Technological Research Council of Turkey) under
two career awards (104E057 and 104E058). The first author was
supported by a TUBITAK BIDEB Fellowship.}
\normalsize Center for Advanced Design Technologies \\
College of Engineering \\
{\normalsize Ko{c} University, Istanbul, Turkey\\
\vspace{0.25cm}
November 19, 2007} \vspace{-0.2cm} }

\begin{document}

\sloppypar
\ssp

\maketitle


\begin{abstract}
\input{abs}
\end{abstract}



\vspace{-0.0cm}
\section{Introduction}
\label{sec:intro}
\input{intro}


\vspace{-0.0cm}
\section{Background}
\label{sec:bg}
\input{bg}


\section{Timing Yield Estimation with ISLE}
\label{sec:method}
\input{methodPre}
\input{method}
\input{methodExp}


\section{Results}
\label{sec:case}
\input{case}

\section{Conclusion}
\label{sec:conc}
\input{conc}

\bibliographystyle{unsrt}

\bibliography{references}

\end{document}

%% file: header.tex
\usepackage{cite}
\usepackage{graphicx}  
\usepackage{psfrag}   
\usepackage{subfigure} 
\usepackage{url}       
\usepackage{amsmath}   
\usepackage{array}

%% file: commands.tex
\newcommand{\ignore}[1]{}

\newcommand{\halmos}{\hskip\textwidth minus\textwidth \EndProof}

\newtheorem{theorem}{Theorem}[section]

\def\EndProof{\hfill \quad \vrule width 1ex height 1ex depth 0pt  }

\renewcommand{\jmath}{{j}}

\newcommand {\beq} {\begin{equation}}
\newcommand {\eeq} {\end{equation}}

\newcommand {\eq}[1] {(\ref{eqn:#1})}

\newcommand {\secc}[1] {Section~\ref{sec:#1}}

\newcommand {\figg}[1] {Figure~\ref{fig:#1}}
\newcommand {\tabb}[1] {Table~\ref{tab:#1}}

\newcommand {\thmm}[1] {Theorem~\ref{thm:#1}}

\newcommand {\lrp}[1]  {{ \left({#1}\right) }}

\newcommand{\Ckt}{\mathcal{C}}
\newcommand{\gate}{g}
\newcommand{\bi}{\begin{itemize}}

\newcommand{\ei}{\end{itemize}}

\newcommand{\Yield}{\textit{Yield}}

\newcommand{\be}{\begin{equation}}
\newcommand{\ee}{\end{equation}}

\newcommand{\gatdelLE}{{d_r^{LE}}}

\newcommand{\refinvdel}{{\tau}}
\newcommand{\refinvdelX}{{\tau(X)}}
\newcommand{\lefac}{{d}}

\newcommand{\paras}{{p}}
\newcommand{\logeffort}{{g}}
\newcommand{\fanout}{{h}}
\newcommand{\aPath}{\pi}
\newcommand{\dpi}[1]{d_\pi}
\newcommand{\xx}{X}

\newcommand{\slev}{\ensuremath{\mathit{SLE.d1}}}
\newcommand{\slevv}{\ensuremath{\mathit{SLE.d2}}}
\newcommand{\parasX}{{p(X)}}
\newcommand{\logeffortX}{{g(X)}}

\newcommand{\GG}{G}
\newcommand{\tGG}{{\tilde{G}}}
\newcommand{\Gn}{G_N}
\newcommand{\gfunc}{g}
\newcommand{\var}{\sigma^2}

\newcommand{\ffunc}{f}

\newcommand{\ftil}{\tilde{f}}
\newcommand{\ind}{I(\timeConstr,\aPP)}
\newcommand{\indi}{I(\timeConstr,\aPP_i)}

\newcommand{\pathDelayLE}[1]{d^{\textit{LE}}_{#1}(\aPP)}

\newcommand{\timeConstr}{T_c}
\newcommand{\lLEeps}{\textit{Loss}^{LE,\epsilon}}
\newcommand{\lLE}{\textit{Loss}^{LE}}
\newcommand{\Loss}{\textit{Loss}}

\newcommand{\pathDelay}[1]{d_{#1}(\aPP)}
\newcommand{\circDelay}{{d_{\mathcal{C}}(\aPP)}}
\newcommand{\circDelayi}{{d_{\mathcal{C}}(\aPP_i)}}
\newcommand{\circDelayLE}{{d_{\mathcal{C}}^{LE}(\aPP)}}
\newcommand{\circDelayLEi}{{d_{\mathcal{C}}^{LE}(\aPP_i)}}
\newcommand{\CritPaths}{{\Pi_{crit}}}
\newcommand{\critPaths}{\CritPaths}

\newcommand{\aProcParam}{X}
\newcommand{\aPP}{\aProcParam}

\newcommand{\pdf}[1]{f({#1})}

\newcommand{\indSLEi}{{I^{LE}(\timeConstr,\aPP_i)}}
\newcommand{\indSLE}{I^{\textit{LE}}(\timeConstr,\aPP)}
\newcommand{\indLEi}{{I^{LE}(\SLEtimeConstr,\aPP_i)}}
\newcommand{\indLE}{I^{\textit{LE}}(\SLEtimeConstr,\aPP)}

\newcommand{\SLEtimeConstr}{\timeConstr^{\epsilon}}

\newcommand{\errMC}{{Error_{MC}}}
\newcommand{\errISLE}{{Error_{ISLE}}}

\newcommand{\gain}{\text{\sf Gain}}
\newcommand{\numMC}{N_{MC}}
\newcommand{\numISLE}{N_{ISLE}}

\newcommand{\alg}{\textsc{IsleExplorer}}
\newcommand{\algg}{\textsc{Explore}}
\newcommand{\MCSimCapacity}{{MCSimCapacity}}

\newcommand{\Leff}{{L_{eff}}}
\newcommand{\Vdd}{{V_{dd}}}
\newcommand{\Vth}{{V_{th}}}
\newcommand{\exper}{{\bf Exp}}
\newcommand{\variii}{\sf ThrPar}
\newcommand{\varii}{\sf TwoPar}
\newcommand{\vari}{\sf OnePar}
\newcommand{\invCh}{{\sf InverterChain}}
\newcommand{\randCh}{{\sf GateChain}}

\ignore{

\begin{figure*}[!htb]
...
\end{figure*}

 \begin{table*}[t]
\begin{equation}
...
\end{equation}
\hrule 
\end{table*}

}


%% file: abs.tex
In the nano era in integrated circuit fabrication technologies, the
performance variability due to statistical process and circuit
parameter variations is becoming more and more significant.
Considerable effort has been expended in the EDA community during the
past several years in trying to cope with the so-called {\em
statistical timing} problem. Most of this effort has been aimed at
generalizing the static timing analyzers to the statistical case.  In
this paper, we take a pragmatic approach in pursuit of making the
Monte Carlo method for timing yield estimation practically feasible.
The Monte Carlo method is widely used as a golden reference in
assessing the accuracy of other timing yield estimation
techniques. However, it is generally believed that it can not be used
in practice for estimating timing yield as it requires too many costly
full circuit simulations for acceptable accuracy. In this paper, we
present a novel approach to constructing an improved Monte Carlo
estimator for timing yield which provides the same accuracy as the
standard Monte Carlo estimator, but at a cost of much fewer full
circuit simulations. This improved estimator is based on a novel
combination of a variance reduction technique, importance sampling,
and a stochastic generalization of the logical effort formalism for
cheap but approximate delay estimation. The results we present
demonstrate that our improved yield estimator achieves the same
accuracy as the standard Monte Carlo estimator at a cost reduction
reaching several orders of magnitude.

%% file: intro.tex
\noindent We address the problem of estimating timing yield for a
circuit under statistical process parameter variations and
environmental fluctuations by proposing a novel and improved Monte
Carlo method based on transistor-level circuit simulations.  In
conventional Monte Carlo yield estimation, a number of samples in the
parameter probability space are generated.  The overall delay of the
circuit for each sample point is determined by performing
transistor-level timing simulations.  An estimator for timing yield is
obtained by considering the fraction of samples for which the timing
constraint is satisfied.  Because of the computational cost of
determining circuit delay for each sample, the number of samples one
has to work with is limited.  This adversely affects the accuracy of
the yield estimator; which has a large error for a small number of
samples. This is a weakness of the conventional Monte Carlo method and
has prevented it from finding widespread use for practical yield
estimation.

The technique we propose aims to improve the accuracy of the yield
estimates obtained from a given number of Monte Carlo
simulations. Alternatively, our improved Monte Carlo estimator
achieves the same accuracy as the standard Monte Carlo estimator, but
at a cost of much fewer number of full circuit simulations.  This is
made possible by using a variance reduction technique called
importance sampling that we combine in a novel manner with a
stochastic generalization of the logical effort formalism originally
proposed by Sutherland~et.~al~\cite{logeffortbook}. 
Logical effort is a method for quickly
estimating and optimizing the path delays in a circuit.  We use the
stochastic logical effort formalism to guide the generation and
selection of sample points in the parameter probability space in a
transistor-level simulation based Monte Carlo method for timing yield
estimation.

Our approach is based on the premise that, given the magnitude of
process parameter variations and the non-linear dependency of gate and
circuit delay on these variations, the only sufficiently reliable and
accurate method for determining circuit delay is detailed,
transistor-level simulation. We believe that sufficient accuracy in
yield estimation can not be obtained even by applying Monte Carlo
simulations at a higher level, e.g. at the block level. Yield
estimation techniques not based on Monte Carlo simulations operate by
propagating probability density functions across the circuit. To make
this process feasible, one is forced to use approximate gate delay
models and delay propagation methods that may be too inaccurate when
process paramater variations are large.  We therefore believe that
accurate determination of timing yield must have circuit simulation as
its basis.  We demonstrate in this paper that Monte Carlo simulation
in conjunction with a novel variance reduction technique can serve as
an accurate yet computationally viable yield estimation method.

In \secc{bg}, we provide background information and
preliminaries on logical effort and its stochastic generalization, the
Monte Carlo method for yield estimation and importance sampling. In
\secc{method}, we present our improved Monte Carlo yield estimator
featuring significant error reduction through the use of the
stochastic logical effort formalism to facilitate importance sampling.
We also provide a precise, comparative error analysis for our proposed
technique and the standard Monte Carlo estimator. Finally in
\secc{case}, we present experimental results on several examples which
demonstrate the efficiency and accuracy of our proposed yield
estimator.


%% file: bg.tex
\noindent
Section~\ref{sec:SLE} provides an overview of the logical effort
approach \cite{logeffortbook}. In Section~\ref{sec:SSLE}, we introduce
two techniques for using logical effort as a method for approximating
circuit delay in the presence of statistical variations.
Section~\ref{sec:mc} reviews the standard Monte Carlo method for
evaluating definite integrals, and Section~\ref{sec:is} presents the
importance sampling technique for variance reduction in Monte Carlo
simulations. Finally, Section~\ref{sec:defn} describes the
preliminaries in applying the standard Monte Carlo and importance
sampling techniques to timing yield estimation.

\subsection{Logical Effort}
\label{sec:SLE}
\noindent
The logical effort formalism is a fast and efficient way of
determining the delay of a path in a digital circuit.  The path delay
is simply the sum of the delays of the gates on the path, and the
delay of a logic gate $r$ is approximated as
\begin{equation}
\gatdelLE = \refinvdel\;\lefac
\label{eqn:gatedelay}
\end{equation}
where $\gatdelLE$ is the absolute delay of a gate measured in seconds,
$\refinvdel$ is the delay of a parasitic-capacitance-free {\em
  reference inverter} driving another identical inverter, and $\lefac$
is the delay of the logic gate expressed in units of $\refinvdel$. The
$\lefac$ factor in \eq{gatedelay} models the gate delay and is given
by
\begin{equation}
\lefac = \lrp{\paras+\logeffort\,\fanout}
\label{eqn:lefac}
\end{equation}
where $\paras$ represents the intrinsic (parasitic) delay,
$\logeffort$ is the logical effort, and $\fanout$ is the electrical
effort or electrical fan-out. Logical effort $\logeffort$ for a logic
gate is defined as the (unitless) ratio of its (per) input capacitance
to that of an inverter that delivers the same output current. Thus,
logical effort $\logeffort$ is a measure of the complexity of a
gate. It depends only on the gate's topology and is independent of the
size and the loading of the gate. Parasitic delay $\paras$ expresses
the intrinsic delay of the gate due to its own internal parasitic
capacitance, and it is largely independent of the sizes of the
transistors in the gate.  Parasitic delay $\paras$, is also expressed
in units of $\refinvdel$. The electrical effort $\fanout$ is the ratio
of the load capacitance of the logic gate to the capacitance of a
particular input~\cite{logeffortbook}.
\subsection{Stochastic Logical Effort}
\label{sec:SSLE}
\noindent
Equations (\ref{eqn:gatedelay}) and (\ref{eqn:lefac}) provide a way
of decomposing the effects of statistical parameter variations on gate
delays.  In a different context,
Sutherland~et.~al~\cite{logeffortbook} analyzed different
semiconductor processes with varying supply voltages, and observed
that almost all of the effect of process parameters and supply voltage
on gate delay is captured by the reference inverter delay
($\refinvdel$ in \eq{gatedelay}), even when the parameters vary over a
large range spanning different fabrication processes.  The logical
effort $\logeffort$ and the unitless parasitic delay $\paras$ of a
gate exhibit relatively little variation with process parameters and
supply voltage.  Exploiting this observation in the context of timing
yield analysis, in \cite{sleTAU2006} a stochastic logical effort (SLE)
model was proposed where the delay of a gate was modeled as
\begin{equation}
\label{eqn:SLE-Eq-v1} \gatdelLE\lrp{\xx} =
\refinvdelX\;\lrp{\paras+\logeffort\,\fanout}
\end{equation}
where $\xx$ is a vector of random variables, each component of which
represents a different statistical circuit or process parameter and
$\refinvdelX$ is the reference inverter delay when the parameters are
given by $\xx$. As is apparent in this equation, in the stochastic
logical effort approximation, all process and environmental variations
are captured by the statistical variable $\refinvdel$ while
$\logeffort$, $\paras$ and therefore $\lefac$ are assumed to be
independent of process parameters. If
only inter-die variations are modeled, statistical parameters on the
chip at all locations are perfectly correlated. In this case, using
the stochastic characterization of $\refinvdel$ for the same reference
inverter for all of the logic gates on the die captures this perfect
statistical correlation among gates. We refer to the
approximation given in \eq{SLE-Eq-v1} as {\em
  first-degree stochastic logical effort} (abbreviated as $\slev$).

In this paper, we also investigate a further refinement of this
approximation described by the following equation
\begin{equation}
\label{eqn:SLE-Eq-v2} \gatdelLE\lrp{\xx} =
\refinvdelX\;\lrp{\parasX+\logeffortX\,\fanout}
\end{equation}
where the dependency of $\paras$ and $\logeffort$ on $\xx$ is also
modeled. We call this model {\em second-degree stochastic logical
  effort} ($\slevv$).  As will become apparent later in the paper,
$\slevv$ is more accurate but computationally more expensive.

In both versions of SLE, in order to compute the
delay of a path $\aPath$ in a circuit, we simply add the delays of the
gates on $\aPath$:
\begin{equation}
\label{eqn:pathDelay1} \pathDelayLE{\aPath} = \sum_{r=1}^k
\gatdelLE\lrp{\xx}
\end{equation}
Here $\gatdelLE\lrp{\xx}$ is the delay of the $r$-th gate on the path
$\aPath$. $\gatdelLE\lrp{\xx}$ is computed by evaluating
\eq{SLE-Eq-v1} for $\slev$ and \eq{SLE-Eq-v2} for $\slevv$. For this
evaluation, a full transistor-level simulation of the whole circuit
containing the logic path is not necessary. However, the values of
$\refinvdelX$ (for both $\slev$ and for $\slevv$), and $\parasX$ and
$\logeffortX$ (for $\slevv$) at a given $\xx$ are needed. For the
results we present in this paper, we compute these at a given $\xx$ by
running transistor-level circuit simulations on small test circuits
which contain only the reference inverter (for $\refinvdelX$) or the
gate under consideration (for $\parasX$ and $\logeffortX$) together
with a proper driver and load circuitry. We envision that the
statistical characterizations (in the form of simple analytical
formulas, a look-up table or a response surface model generated by
running circuit simulations on appropriate test circuits) for these
quantities could become part of the characterizations supplied with a
standard-cell library. In this case, no circuit simulations will be
needed when evaluating the SLE delay formulas for circuits that are
built using gates from such a pre-characterized library.
\subsection{The Monte Carlo Method}
\label{sec:mc}
\noindent Monte Carlo (MC) techniques can be used
to estimate the value of a definite, finite-dimensional integral
of the form
\begin{equation}
 \GG = \int_\Omega \gfunc(\xx) \ffunc(\xx) d\xx
\label{eqn:vanil} \end{equation}
\noindent where $\Omega$ is a finite domain and $\ffunc(\xx)$ is a
probability density function (PDF) over $\xx$, i.e., $\ffunc(\xx) \geq
0$ for all $\xx$ and $\int_\Omega \ffunc(\xx) d\xx = 1$.
MC estimation for the value of $\GG$ is accomplished by
drawing a set of independent samples $\xx_1, \xx_2, ..., \xx_N$ from
$\ffunc(\xx)$ and by using

\begin{equation}
G_N = (1/N) \sum_{i=1}^N \gfunc(\xx_i)
\label{vanil-estim} \end{equation}
The estimator $\Gn$ above is itself a random variable.  Its mean is
equal to the integral $\GG$ that it is trying to estimate, i.e.,
$E(\Gn)=\GG$, making it an unbiased estimator.  The variance of $\Gn$
is $Var(\Gn)= {\var}/{N}$, where $\var$ is the variance of
the random variable $\gfunc(\xx)$ given by
\begin{equation}
\var = \int_\Omega  \gfunc^2(\xx) \ffunc(\xx) d\xx - \GG^2
\label{eqn:sigma-sqr}
\end{equation}
The standard deviation of $\Gn$ can be used to assess its
accuracy in estimating $\GG$. If $N$ is sufficiently large, due to the
Central Limit Theorem, $\tfrac{\Gn-\GG}{\sigma/\sqrt{N}}$ has an
approximate standard normal ($N(0,1)$) distribution. Hence,
\begin{equation} P(\GG -
  1.96\tfrac{\sigma}{\sqrt{N}}\leq \Gn \leq \GG +
  1.96\tfrac{\sigma}{\sqrt{N}}) =
  0.95 \label{eqn:normal-error}
\end{equation}
where $P$ is the probability measure. The equation above means that
$\Gn$ will be in the interval $[\GG - 1.96\tfrac{\sigma}{\sqrt{N}},
\GG + 1.96\tfrac{\sigma}{\sqrt{N}}]$ with 95$\%$ confidence. Thus, one
can use the error measure
\begin{equation}
|Error| \approx {\tfrac {2 \sigma}{\sqrt{N}}}
\label{eqn:error-est} \end{equation}
in order to assess the accuracy of the estimator.

Several techniques exist for improving the accuracy of MC evaluation
of finite integrals. In these techniques, one tries to construct an
estimator with a reduced variance for a given, fixed number of
samples, or equivalently, the improved estimator provides the same
accuracy as the standard MC estimator but with considerably fewer
number of samples. This is desirable because computing the value of
$g(X_i)$ is typically computationally or otherwise costly.
\subsection{Importance Sampling}
\label{sec:is}
\noindent
One MC variance reduction technique is importance sampling
(IS)~\cite{KalosWhitlock86,StratMC}. IS improves upon the standard MC
approach described above by drawing samples for $\xx$ from another
distribution $\ftil$.  $\GG$ in \eq{vanil} is first rewritten as below
\begin{equation} \GG = \int_\Omega
  \left(
    \tfrac{\gfunc(\xx) \ffunc(\xx)}{\ftil(\xx)}
  \right)
  \ftil(\xx) d\xx
  \label{eqn:impSampl}
\end{equation}
If $\xx_1, \xx_2, ..., \xx_N$ are drawn from $\ftil$ instead of
$\ffunc$, the improved estimator $\tGG_N$ takes the form
\begin{equation}
\tGG_N = \frac{1}{N} \sum_{i=1}^N  \gfunc(\xx_i)\tfrac{\ffunc(\xx_i)}{\ftil(\xx_i)}
\label{eqn:estim}
\end{equation}
where the weighting factor ${\ffunc(\xx_i)}/{\ftil(\xx_i)}$ has been
used in order to compensate for the use of samples drawn from the
biased distribution $\ftil$.  In order for the improved estimator
above to be well-defined and unbiased, $\ftil(\xx_i)$ must be nonzero
for every $\xx_i$ for which $\ffunc(\xx_i) \gfunc(\xx_i)$ is
nonzero. We refer to this as the {\em safety requirement}. The ideal
choice for the biasing distribution $\ftil$ is
 \begin{equation}
\ftil_{ideal}(\xx) =
  \tfrac{\gfunc(\xx)\,\ffunc(\xx)}{G}
\end{equation}
 which results in an exact
  estimator with zero variance with a single sample! However,
  $\ftil_{ideal}$ obviously can not be used in practice since the
  value of $G$ is not known a priori. Instead, a practically
  realizable $\ftil$ that resembles $\ftil_{ideal}$ is used. The
  key in using IS in practical problems is the determination of
  an effective biasing distribution that results in significant
  variance reduction. We have identified one such biasing distribution
  by exploiting the SLE formalism that we use to
  construct an efficient and accurate estimator for the timing
  yield of digital circuits. This distribution will be described in
  Section~\ref{sec:method}.
\subsection{Monte Carlo Estimation of Timing Yield}
\label{sec:defn}
\noindent
A path $\pi$ in a circuit
$\Ckt$ is a sequence of gates $\gate_0, \gate_1, \gate_2,
...,\gate_n$ where $\gate_0$'s inputs are primary inputs of the
circuit, and $\gate_n$'s output is a primary output of the circuit.
Given a circuit and values for the statistical parameters, a path is
said to be {\em critical} if (i) it is sensitizable, and (ii) its
delay is as large as the delays of other sensitizable paths. A
path $\aPath$ is said to be {\em statistically critical} if it is
a critical path of $\Ckt$ for some possible assignment to process
parameters. We denote by $\critPaths$ the set of statistically
critical paths. Then, the delay of a circuit is computed using
\begin{equation}
\circDelay = max_{\aPath \in \CritPaths} \:\: \pathDelay{\aPath}
\label{eqn:dc}
\end{equation}
where $\circDelay$ is the delay of the circuit and
$\pathDelay{\aPath}$ is the delay of path $\aPath$ when the
circuit and process parameters are given by $\aPP$.

A target delay $\timeConstr$ is specified for the circuit. Given a
PDF $\ffunc(\xx)$ for the statistical
parameters, we would like to compute the fraction of circuits that
satisfy $\circDelay \leq \timeConstr$, i.e., the {\em timing yield} of
the circuit.  We define an indicator random variable $\ind$ for the
entire circuit as follows: $\ind = 1$ if the circuit delay exceeds the
target, i.e., $\circDelay > \timeConstr$, and $\ind = 0$ otherwise. We
then define the {\em timing loss} or simply {\em loss} with

\begin{equation}
\Loss = 1 - \Yield = \int \ind \: \pdf{\aPP} \: d \aPP
\label{eqn:yield}
\end{equation}
as the mean of the indicator random variable $\ind$ over the PDF $\pdf{\aPP}$.
Evaluation of the integral above is the timing
yield (loss) estimation problem addressed in this paper.

In a straightforward application of the MC method to loss estimation,
one would draw samples $\xx_1, \xx_2, ..., \xx_N$ from the statistical
parameter space according to the PDF $\ffunc(\xx)$ and
construct the {loss} estimator
\begin{equation} \Loss_N = \frac{1}{N} \sum_{i=1}^N
\indi
\label{eqn:LossN-estim}
\end{equation}
With the MC method, full circuit simulations (transistor-level SPICE
simulations of the whole circuit containing the paths under
consideration) must be performed for each sample point, $\aPP_i$, in
order to compute $\circDelayi$ and determine whether $\indi=1$ or $0$.
The MC method is widely used as a golden reference in the literature
in assessing the accuracy and efficiency of timing yield estimation
techniques.  However, it is generally believed that it can not be used
in practice for estimating timing yield as it requires too many costly
full circuit simulations for acceptable accuracy, even though there
are some arguments to the contrary~\cite{SchefferTAU2004}. In the rest of
this paper, the {loss} estimator in (\ref{eqn:LossN-estim}) is
referred to as the standard MC (STD-MC) estimator.

{Loss} can also be estimated based on the SLE formalism,
without performing any full circuit simulations.  The delay of a
circuit can be computed analytically based on the SLE formalism as
follows
\begin{equation}
  \circDelayLE = max_{\aPath \in \CritPaths} \:\:
  \pathDelayLE{\aPath}
  \label{eqn:dcLE}
\end{equation}
where $\pathDelayLE{\aPath}$ is evaluated using the SLE formula in
(\ref{eqn:pathDelay1}) and using $\slev$ or $\slevv$.  We define a new
indicator random variable $\indSLE$, which takes the value $1$ if the
delay of a circuit computed analytically using the SLE equations
exceeds the target delay $\timeConstr$, i.e., $\indSLEi$ is $1$ if
$\circDelayLEi > \timeConstr$, and $0$ otherwise. The {loss} estimator
based on this new indicator variable takes the form
\begin{equation}
\lLE_N = \frac{1}{N} \sum_{i=1}^N \indSLEi
\label{eqn:lossLE-mc}
\end{equation}
In computing $\lLE_N$ above, no full circuit simulations are
performed. Only simple evaluations of the SLE delay formulas are
needed, based on pre-characterizations of
$\refinvdelX$, $\parasX$ and $\logeffortX$) in \eq{SLE-Eq-v1} and
\eq{SLE-Eq-v2}. In contrast, the {loss} estimator in
(\ref{eqn:LossN-estim}) requires $N$ full circuit simulations, one
for every sample. The loss estimator in (\ref{eqn:lossLE-mc}) will be
referred to as the {SLE-MC estimator} in the rest of this paper.

The estimation of {loss} based on the STD-MC estimator in
(\ref{eqn:LossN-estim}) will obviously be more accurate than the one
based on the SLE-MC estimator in (\ref{eqn:lossLE-mc}), but much more
costly. We use the cheap SLE-MC estimator not by itself for yield
estimation, by in a novel approach to constructing an IS-based {loss}
estimator with reduced variance.  This approach is called ISLE
(Importance Sampling based on Stochastic Logical Effort) and provides
the same accuracy as the STD-MC estimator but at a cost of much fewer
number of full circuit simulations.

%% file: methodPre.tex
\noindent The biasing distribution $\ftil(\xx)$ used in ISLE is
\begin{equation}
\ftil(\xx) = \frac{\indLE \ffunc(\xx)}{\lLEeps}
\label{eqn:ftil-SLE}
\end{equation}
This biasing distribution serves as a good approximation to the
ideal but practically unrealizable biasing distribution for
importance sampling, $\ftil(\xx) = \tfrac{\ind \ffunc(\xx)}{\Loss}
\label{eqn:ideal-ftil}$. In \eq{ftil-SLE} above,
$\SLEtimeConstr=(1-\epsilon)\timeConstr$ and $\lLEeps$ is the loss
computed by the SLE-MC estimator in (\ref{eqn:lossLE-mc}) with the
target delay set to $\SLEtimeConstr$ instead of $\timeConstr$,
where $\epsilon$ is a margin parameter which we explain below. The
ISLE loss estimator is then constructed as follows
\begin{equation}
\Loss_N^{ISLE} = \frac{1}{N} \sum_{i=1}^N \indi \tfrac{\ffunc(\xx_i)}{\ftil(\xx_i)}
\label{eqn:IS-estim}
\end{equation}
where the sample points $\xx_i$ must be drawn from $\ftil(\xx)$
 in (\ref{eqn:ftil-SLE}) instead of $\ffunc(\xx)$. The margin
parameter $\epsilon$ was introduced above in order to guarantee that
$\ftil(\xx_i)$ is nonzero everywhere $\indi \ffunc(\xx_i)$ is
nonzero, i.e., $\indLEi$ must take the value $1$ everywhere $\indi$ is
$1$.  The margin parameter $\epsilon$ must be large enough so that the
indicator variables never assume the values $\indLEi=0$ (the timing
constraint $\SLEtimeConstr$ is satisfied according to SLE) and
$\indi=1$ (the actual circuit fails to satisfy the timing constraint)
for any of the sample points.  An automated and adaptive algorithm for
the determination of the smallest value for the margin parameter
$\epsilon$ will be described in Section~\ref{sec:margin-IS}

Substituting the biasing distribution $\ftil$ in
(\ref{eqn:ftil-SLE}) into (\ref{eqn:IS-estim}), and
performing some simplications based on the fact that $\indLEi$ takes the
value $1$ for all samples drawn from $\ftil(\xx)$, we arrive at
\begin{equation}
\Loss_N^{ISLE} = \frac{\lLEeps}{N} \;\;\sum_{i=1}^N {\indi}
\label{eqn:IS-estim2}
\end{equation}
where, as in \eq{IS-estim}, the samples $\xx_i$ are
drawn from $\ftil(\xx)$ in (\ref{eqn:ftil-SLE}).

In order to draw a sample from $\ftil(\xx)$ in (\ref{eqn:ftil-SLE}),
we first draw a sample from $\ffunc(\xx)$. We keep the sample if
$\indLEi$ evaluates to 1 at the sample point and discard it otherwise.
The evaluation of $\indLEi$ is done using the analytical SLE formulas.
Each kept sample constitutes one of the $\xx_i$ in \eq{IS-estim2}.
$\Loss_N^{ISLE}$ is then computed by
determining whether $\indi=1$ for each such kept sample, i.e., by carrying
out a full circuit-level simulation at $\xx_i$.

A key benefit of the ISLE approach is that circuit-level simulations
are avoided for discarded samples, i.e., when $\xx_i$ results in an SLE
circuit delay estimate smaller than $\SLEtimeConstr$. The improvement
brought about by ISLE, however, goes significantly beyond this. For
the same number of samples $N$, the ISLE estimator in \eq{IS-estim2}
provides a much more accurate (with significantly reduced variance)
loss estimate than the STD-MC estimator in \eq{LossN-estim}. Were it
possible to use the ideal biasing function $\ftil_{ideal}$, a
zero-variance estimator would have been obtained with a single sample.
The ISLE approach makes it possible to explore the space between
standard MC and this ideal. Using an $\ftil$ that approximates
$\ftil_{ideal}$ as closely as possible, ISLE both reduces the number
of full circuit simulations required {\em and} improves upon standard
MC in the estimator accuracy achieved for the same number of full
circuit simulations. The next section makes this discussion more
precise.


%% file: method.tex
\subsection{Theoretical Gain: Quantifying Variance Reduction due to ISLE}
\label{sec:theo-gain}
\noindent
The error of an estimator is the deviance of the estimator's
result from the actual loss as explained in
Section~\ref{sec:mc} for a general estimator. In this
section, the errors of the STD-MC and ISLE estimators are
derived and the results are compared.
\begin{theorem}
  The error of the STD-MC estimator in \eq{LossN-estim} obtained with
  $N$ full-circuit simulations is
\begin{equation}
\errMC = \tfrac {2\sqrt {\Loss.\Yield}}{\sqrt{N}}
\label{eqn:error-MC}
\end{equation}
with more than 95\% confidence.
\end{theorem}
\begin{proof}
  By (\ref{eqn:error-est}), the error of the STD-MC estimator for
  loss using $N$ full-circuit simulations is
  $2\sigma/\sqrt{N}$ where $\sigma^2$ is the variance
  of the indicator random variable $\ind$ with PDF $\ffunc(\xx)$.
The mean of $\ind$ is equal to the actual timing loss.
$\sigma^2$ is computed as
\begin{equation}
\sigma^2 = \int_\Omega \ind^2 \ffunc(\xx) d\xx - Loss^2
\label{eqn:MC-sigma} \end{equation}
$\ind$ is either 1 or 0, thus, $\ind$ = $\ind^2$.
Eqn \eq{MC-sigma} becomes
\begin{equation}
\sigma^2 = \Loss - \Loss^2 = \Loss (1 - \Loss) = \Loss.\Yield
\label{eqn:MC-sigma2}\end{equation} The error of the STD-MC
estimator is thus given by \eq{error-MC}.
\end{proof}
\begin{theorem}
\label{thm:thm2}
The error of the ISLE estimator for loss when $N$ full circuit
simulations are performed is
\begin{equation}
\errISLE = \left. {2\sqrt{\Loss.(\lLEeps - \Loss)}} \right/ {\sqrt{N}}
\label{eqn:error-IS-SLE}
\end{equation}
with more than 95\% confidence.
\end{theorem}
\begin{proof}
  By (\ref{eqn:error-est}), the error of the ISLE estimator for
  loss using $N$ full-circuit simulations is
  $2\tilde{\sigma}/\sqrt{N}$ where $\tilde{\sigma}^2$ is the variance
  of the random variable $\tfrac{\ind \ffunc(\xx)}{\ftil(\xx)}$ with PDF $\ftil(\xx)$.
The mean of this random variable  is equal to the actual timing loss.
$\tilde{\sigma}^2$ is computed as
\begin{equation}
\tilde{\sigma}^2 = \int_\Omega {(\tfrac{\ind \ffunc(\xx)}{\ftil(\xx)})}^2 \ftil(\xx) d\xx - Loss^2
\label{eqn:IS-SLE-sigma1}
\end{equation}
Substituting $\ftil(\xx)$ from (\ref{eqn:ftil-SLE}) and using
the fact that $\ind^2 = \ind$ we obtain
\begin{equation}
\tilde{\sigma}^2 = \int_\theta \frac{\ind \ffunc^2(\xx)}
                            {\frac{\indLE \ffunc(\xx)}
                                  {\lLEeps}
                            }
             d\xx - Loss^2
\label{eqn:IS-SLE-sigma2}
\end{equation}
$\theta$ denotes the subregion of $\Omega$ in which $\ftil(\xx)$ is non-zero.
From \eq{ftil-SLE}, $\ftil(\xx)$ is zero when $\indLE$ is zero (and thus $\ind=0$, if the margin $\epsilon$ is chosen properly). When $\ftil(\xx)$ is non-zero,
$\indLE=1$. Thus
\begin{equation}
\tilde{\sigma}^2 = \lLEeps \int_\theta \ind \ffunc(\xx) d\xx - Loss^2 =
\Loss . (\lLEeps - Loss)
\label{eqn:IS-SLE-sigma3}
\end{equation}
The error of the ISLE estimator is thus given by
\eq{error-IS-SLE}.
\end{proof}

If the same number of full circuit simulations $N$ is used with
both methods, then the ratio of the errors of the estimators is given by
\begin{equation}
\textit{Error Ratio} = \frac {\errMC}{\errISLE} = \sqrt {\tfrac
{\Yield}{(\lLEeps-\Loss)}}
\label{eqn:error-ratio}
\end{equation}
Alternatively, suppose a bound on the allowable estimation error
is given.  The ratio of the number of full circuit
simulations required by the two approaches to achieve this same error bound
is given by
\begin{equation}
\gain = \frac {\numMC}{\numISLE} = \frac
{\sigma^2}{\tilde{\sigma}^2}= \frac {\Yield}{(\lLEeps- Loss)}
\label{eqn:gain1}
\end{equation}
As is apparent from (\ref{eqn:error-ratio}) and (\ref{eqn:gain1}), as $\lLEeps$ approaches the real loss $Loss$, the improvement that ISLE offers over STD-MC increases.

In the proof of \thmm{thm2} above for the error of the ISLE estimator,
$\lLEeps$ was assumed to be a known deterministic quantity.  However,
$\lLEeps$ is not determined analytically. $\lLEeps$ is a random
variable and is estimated using the SLE-MC estimator in
\eq{lossLE-mc}. The variance of this random variable decreases
proportionally to the number of samples used in the SLE-MC
estimator. In order for the error result for the ISLE estimator in
\eq{error-IS-SLE} to be valid, the estimation of $\lLEeps$ must be
performed by using a large enough number of samples in \eq{lossLE-mc}
so that it has negligible variance. This would validate its treatment
as a deterministic quantity in the derivation of the error for the
ISLE estimator. The use of a large number of samples in the SLE-MC
estimator in \eq{lossLE-mc} is easily affordable, because no full
circuit simulations are performed, only simple evaluations of the SLE
delay formulas are needed. The results we present later show that the
theoretical error expressions derived here are in excellent agreement
with experimental data.


%% file: methodExp.tex
\begin{figure}[tb]
\centering \ForceWidth{2in}\centerline{\BoxedEPSF{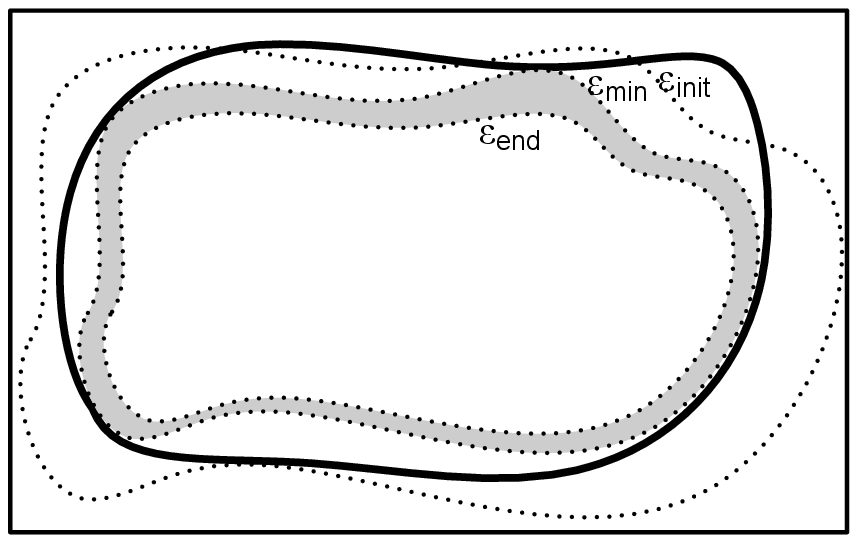}}
\caption{Iterations of the $\alg$ algorithm for determining
the margin $\epsilon$.} \label{fig:golden} \vspace{-0.30cm}
\end{figure}

\begin{algorithm}
\caption{\small \alg($MCSimCapacity$, $ExpectedMaxLoss$,$\timeConstr$)}
\label{alg1}
\algsetup{indent=1em,linenosize=\small, linenodelimiter=.}
\begin{algorithmic}[1]
\small
\STATE $NumFSamples \leftarrow$ $\lceil \MCSimCapacity \times 1/LossBound \rceil$
\STATE Draw $NumFSamples$ sample points $\{\xx_1, \xx_2, \xx_3,..., \xx_{NumFSamples}\}$ from $\ffunc(\xx)$
\FOR{$i=1$ to $NumFSamples$}
\STATE $X_i.color \leftarrow BLACK$
\ENDFOR
\STATE $PointsInMargin \leftarrow 0$,\;
$\epsilon \leftarrow \epsilon_{init}$,\;
$\epsilon{\textit{-step}} \leftarrow 0.02 \textit{ picoseconds}$
\STATE $MCLossCount \leftarrow 0$,\;
$WhitePoints \leftarrow 0$,\;
\WHILE{($PointsInMargin \leq SafetyLimit$)}
\STATE \algg\; ($\timeConstr$, $\epsilon$, $NumFSamples$)
\IF{$PointsInMargin = 0$}
 \IF {(NewWhitePointsDiscovered)}
  \STATE $\epsilon_{min} \leftarrow \epsilon$
  \STATE $LossPointsAt\epsilon_{min} \leftarrow MCLossCount$
  \STATE $WhitePointsAt\epsilon_{min} \leftarrow WhitePoints$
 \ENDIF
\ENDIF
\STATE $\epsilon \leftarrow \epsilon + \epsilon{\textit{-step}}$
\ENDWHILE
\STATE return $(LossPointsAt\epsilon_{min}/WhitePointsAt\epsilon_{min}) \times \lLEeps$
\end{algorithmic}
\end{algorithm}

\begin{algorithm}
\caption{\small \algg\; ($\timeConstr,\; \epsilon,\; NumFSamples$)}
\label{alg2}
\algsetup{indent=1em,linenosize=\small, linenodelimiter=.}
\begin{algorithmic}[1]
\small
\STATE $\SLEtimeConstr \leftarrow \timeConstr - \epsilon$
\STATE $NewWhitePointsDiscovered \leftarrow \mathtt{false}$
\FOR{$i=1$ to $NumFSamples$}
  \IF {$X_i.color=BLACK$}
    \STATE \textbf{Compute} $I^{LE}(\SLEtimeConstr,\,X_i)$
    \IF {$I^{LE}(\SLEtimeConstr,\,X_i)=1$}
      \STATE $X_i.color \leftarrow WHITE$
      \STATE $WhitePoints \leftarrow WhitePoints + 1$
      \STATE $NewWhitePointsDiscovered \leftarrow \mathtt{true}$
      \STATE \textbf{Compute} $I(\timeConstr,\,X_i)\;$ // Full circuit simulation
      \IF {$I(\timeConstr,\,X_i)=1$}
         \STATE $MCLossCount \leftarrow MCLossCount + 1$
         \STATE $PointsInMargin \leftarrow 0$
      \ELSE
         \STATE $PointsInMargin \leftarrow PointsInMargin+1$
      \ENDIF
    \ENDIF
  \ENDIF
\ENDFOR
\end{algorithmic}
\end{algorithm}

\subsection{\alg: The Margin Determination Algorithm}
\label{sec:margin-IS}
\noindent
The determination of an appropriate value of $\epsilon$, the timing
margin used by ISLE, is essential for the correctness, accuracy, and
efficiency of the technique. On the one hand, $\epsilon$ must be large
enough to satisfy the correctness constraint that for every value of
$\xx$ that $\ffunc(\xx).\ind$ is non-zero, $\ftil(\xx)$ is also
non-zero. Since, $\ftil(\xx)$ is proportional to $\ffunc(\xx)$ and
$\indLE$, this translates to the safety requirement that $\indi = 1
\Rightarrow \indLEi = 1$.  On the other hand, as can be seen in
(\ref{eqn:error-IS-SLE}) and (\ref{eqn:gain1}), the closer $\lLEeps$
is to $Loss$, the more accurate the ISLE estimator becomes and the
more improvement it achieves over standard MC. Making $\lLEeps$ close
to $Loss$ requires that $\epsilon$ be kept small. Thus, to make ISLE
accurate and efficient while preserving correctness, we must make
$\epsilon$ as small as possible without violating the safety
requirement for any $\xx_i$.

If SLE were a perfect approximation, a margin of $\epsilon = 0$
would satisfy the requirements above. However, as also
demonstrated by our experimental results, stochastic SLE is
inaccurate to an extent that is circuit and process parameter
dependent. Therefore, the $\epsilon$ margin must be determined
separately for each different circuit and it must be checked that
the resulting $\epsilon$ satisfies $\indi = 1 \Rightarrow \indLE =
1$.

This section presents $\alg$, an iterative heuristic algorithm for determining
$\epsilon$ (Algorithm~\ref{alg1}) and estimating $\Loss$. $\alg$
interleaves steps of incrementing $\epsilon$ and performing a number
of full circuit simulations required to compute the ISLE
estimator. When $\alg$ terminates, a correct value of $\epsilon$ is
determined and all of the full circuit simulations required for computing
the estimator $\Loss_N^{ISLE}$ (\ref{eqn:IS-estim2}) have
been performed. $\alg$ runs only
a fixed number ($SafetyLimit$) of additional full circuit simulations
beyond those needed for $\Loss_N^{ISLE}$ in order to ensure that the
margin value $\epsilon$ used is correct. As will become apparent
below, the cost of the full circuit simulations are the dominant
factor in the computational cost of $\alg$. Therefore, the
computational cost of adaptively determining the margin parameter is a
fixed number $SafetyLimit$ full circuit simulations.

The intuition behind the operation of $\alg$ is illustrated in
Figure~\ref{fig:golden}. The solid rectangle represents the
two-dimensional parameter space. Every possible point in the
rectangle corresponds a unique valuation of the parameters $\xx$.
The solid curve in Figure~\ref{fig:golden} consists of the points
$\xx$ for which the delay of the circuit is exactly $T_c$. Outside
the solid curve, circuit delay exceeds $T_c$, i.e., $\ind = 1$.
Each dotted curve consists of points $\xx$ for which $\circDelayLE
= T_c - \epsilon$ for a particular value of $\epsilon$.

$\alg$ considers $NumFSamples$ samples generated from $\ffunc(\xx)$
and computes $\epsilon$ based on data it collects on these samples.
$\alg$ starts exploration with a negative initial value for the margin
$\epsilon_{init}$, gradually increases $\epsilon$, ends exploration at
$\epsilon_{end}$ and determines $\epsilon_{min}$ in the process.
$\epsilon_{min}$ is the smallest margin $\alg$ can detect for which it
can verify that the $\circDelayLE = T_c - \epsilon_{min}$ curve lies
completely inside the $\circDelay = T_c$ curve. $\epsilon_{min}$ is
the value of $\epsilon$ used by $\alg$ for computing the ISLE
estimator for timing loss (\ref{eqn:IS-estim2}). All circuit
simulations required to compute the summation in (\ref{eqn:IS-estim2})
have already been performed when the $\epsilon$ exploration is
completed. At that point, to arrive at the value of the estimator
$Loss_N^{ISLE}$, all $\alg$ needs is an estimate for the value of
${\textit{Loss}^{LE,\epsilon_{min}}}$ that is computed using the
SLE-MC estimator in \eq{lossLE-mc} as explained before. The
computational cost of $Loss^{LE}$ determination is unavoidable with
ISLE and is not due to the adaptive determination of
$\epsilon$.

At each iteration, $\alg$ increases the margin $\epsilon$ by
$\epsilon\textit{-step}$. It then investigates (using the
\textsc{Explore} subroutine in Algorithm~\ref{alg2}) the samples that
fall between $\circDelayLE = T_c -\epsilon$ and $\circDelayLE = T_c
-\lrp{\epsilon +\epsilon\textit{-step}}$ and determines for each such
sample $\xx_i$ whether $\indi = 1$ is satisfied. The iterations
continue until a value of margin $\epsilon_{end}$ is reached for which
the number of samples $\xx_i$ that fall in a safety band defined by
$I^{\textit{LE}}(T_c - \epsilon_{end},\aPP_i) = 1$ and
$I^{\textit{LE}}(T_c - \epsilon_{min},\aPP_i) = 0$ (also $\indi = 0$)
reaches $SafetyLimit$, a user-given parameter.

$\alg$ uses the colors white and black to mark the status of
samples $\xx_i$ generated from $\ffunc$. If $\xx_i.color = Black$,
this indicates that a full circuit simulation has not been run for
$\xx_i$. This is because for the values of the margin $\epsilon$
explored so far, $I^{LE}(\SLEtimeConstr,\,X_i)$ was found to be $0$.  If
$\xx_i.color = White$, this indicates that an SLE timing estimation
and a full circuit simulation for $\xx_i$ has been performed and it
has been determined whether the safety requirement is satisfied for
$\xx_i$ . White points do not need to be revisited when the value of
$\epsilon$ increases, since the value of $\indi$ and $\indLEi$ do not
change afterwards.

$\alg$ tries to obtain as accurate a delay estimate as possible
while limiting the number of full circuit simulations to about
$MCSimCapacity$, a parameter provided by the user. These $\approx
MCSimCapacity$ samples are chosen among $NumFSamples$ samples
generated from the distribution $\ffunc(\xx)$. The user also
provides a rough estimate for an upper bound on the loss, $0 \leq
ExpectedMaxLoss \leq 1$. From among $NumFSamples$ samples, we
expect to run full circuit simulations for about $ExpectedMaxLoss
\times NumFSamples$. Therefore, the algorithm selects
$NumFSamples$ to be $(1/ExpectedMaxLoss).MCSimCapacity$.

It should be noted that $\alg$ is a heuristic algorithm, and, as such,
does not formally guarantee that the safety requirement is satisfied
for all samples $\xx$. In order to keep the computational cost
reasonable, instead of checking that the safety requirement is
satisfied for all $NumFSamples$ samples $\xx_i$ (since this would
require $NumFSamples$ full circuit simulations) $\alg$ considers
margins larger than the minimum satisfactory $\epsilon_{min}$ and
makes sure that for $SafetyLimit$ samples $\xx_i$ that satisfy $T_c -
\epsilon_{min} \leq \circDelayLEi \leq T_c - \epsilon_{end}$ (the
points in the safety band) the safety requirement is not violated.
This is done in order to build further confidence that the
$\epsilon_{min}$ value arrived at is valid. In future work, we plan to
investigate techniques that can formally ensure that the safety
requirement is satisfied for importance sampling using the SLE
approximation.


%% file: case.tex
\begin{figure}
\centering \ForceWidth{3.5in}
\centerline{\BoxedEPSF{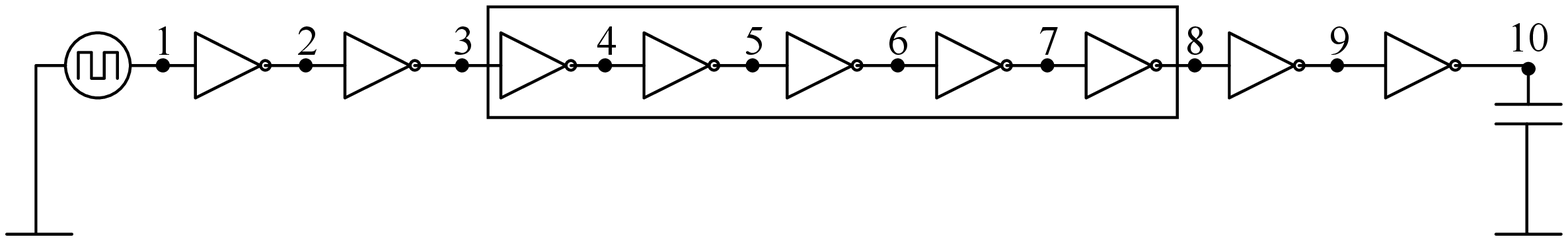}} \vspace{-0.8cm}
\caption{Test Circuit 1: Inverter Chain} \label{fig:TestCrct1}
\end{figure}

\begin{figure}
\centering \ForceWidth{3.5in}
\centerline{\BoxedEPSF{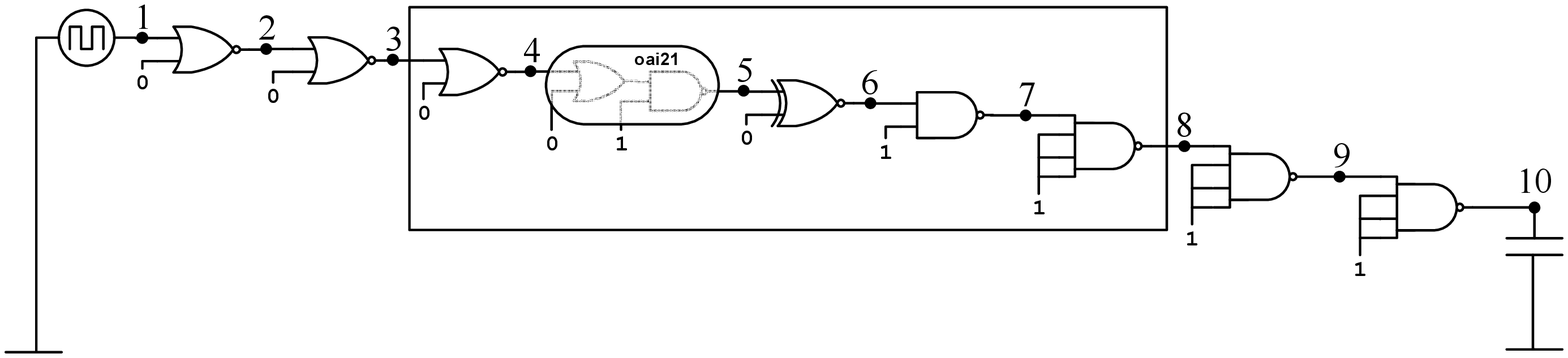}} \vspace{-0.1cm}
\caption{Test Circuit 2: Gate Chain} \label{fig:TestCrct2}
\end{figure}

\subsection{Experimental Setup}
\label{sec:case-exp}
\noindent
We first explain the technical issues related to our experimental
setup in order to help interpret our results better.
We present results on two test circuits, {\invCh} and
  {\randCh} shown in Figures~\ref{fig:TestCrct1} and \ref{fig:TestCrct2}.  For both circuits, the timing loss is
  computed by comparing the delay between nodes 3 and 8 with the
  timing constraint. The precursor gates between nodes 1 and 3, and
  the postcursor gates between nodes 8 and 10 are placed in the
  circuit in order to realize a typical driver and load for the logic
  path under consideration.  The gates used in these circuits are from
  Graham Petley's $0.13\mu$ library version 8.1~\cite{vlsitech}.

We consider three statistically varying process and circuit parameters~\cite{RabaeyELECTRON2003}:
\begin{itemize}
\item Effective channel length ${\Leff}$ with a $3\sigma/\mu$ ratio of 15$\%$.
\item Supply voltage $V_{dd}$ with a $3\sigma/\mu$ ratio of 10$\%$.
\item Threshold voltage $\Vth$  with a $3\sigma/\mu$ ratio of 10$\%$.
\end{itemize}
These parameters are assumed to have Gaussian distributions, and are
considered independent. We create three sets of statistical parameters
from the above:
\begin{itemize}
\item {\vari}: A one-parameter set consisting of ${\Leff}$.
\item {\varii}: A two-parameter set consisting of $\Leff$ and $V_{dd}$.
\item {\variii}: A three-parameter set consisting of $\Leff$, $\Vdd$
  and $\Vth$.
\end{itemize}
For the results we report in this paper, we consider only inter-die
correlations. In other words, the statistical parameters for all of
the transistors in the circuit are fully correlated, and the variation
in the parameters is location and transistor independent.

In order to empirically measure the error in the loss estimates
obtained by the standard MC (STD-MC) estimator and our ISLE estimator,
we perform 50 independent repetitions of the same experiment run. In
our graphs and tables, {\em experiment numbers} $1$ through $50$ refer
to these different runs.  In each independent run, we compute the loss
estimates using $1000$ separate samples generated from $f$ in the
parameter space. These $50$ independent runs constitute samples of the
loss estimator, and the variance and error of the loss estimator is
computed over these $50$ samples.  For the STD-MC estimator,
transistor-level circuit simulations are performed at every one of
these 1000 sample points. For the ISLE estimator, a reduced number of
simulations are performed since most of the samples are discarded
based on the evaluation of the SLE equations.  The number of circuit
simulations that are run for the ISLE estimator may be different in
the 50 runs. In our tables and graphs, we report the average number of
simulations over the 50 runs.

The ${\lLEeps}$ value that is needed for computing the ISLE
  estimator in \eq{ftil-SLE} and \eq{IS-estim2} is computed using
  the SLE-MC estimator in \eq{lossLE-mc} using all of the 50000 sample
  points generated during the 50 runs.

We report and compare loss and error results for three estimators:
\begin{itemize}
\item the standard MC estimator: STD-MC,
\item the ISLE estimator based on {\slev}, and
\item the ISLE estimator based on {\slevv}.
\end{itemize}
\noindent We report results for six different experiment
configurations (combinations of a test circuit and a parameter set) in`
the next two sections. We use the notation {\sf Exp(TestCircuit,
  ParameterSet)} to denote an experiment that is run on the {\sf
  TestCircuit} (which can be {\invCh} or {\randCh}) with the {\sf
  ParameterSet} (which can be {\vari}, {\varii} or {\variii}).

\subsection{The Accuracy and Efficiency of the ISLE Estimator}
\noindent
We illustrate the performance and operation of our ISLE timing
loss estimator (based on $\slevv$) by providing four graphs that were generated
from one selected experiment, {\sf Exp({\randCh},{\variii})}. We are
not able to provide similar graphs for the other experiments due to
space constraints, but we present detailed performance results in the
next section on all of the six experiments we have run.

\begin{figure}
\centering \ForceWidth{2.5in}
\centerline{\BoxedEPSF{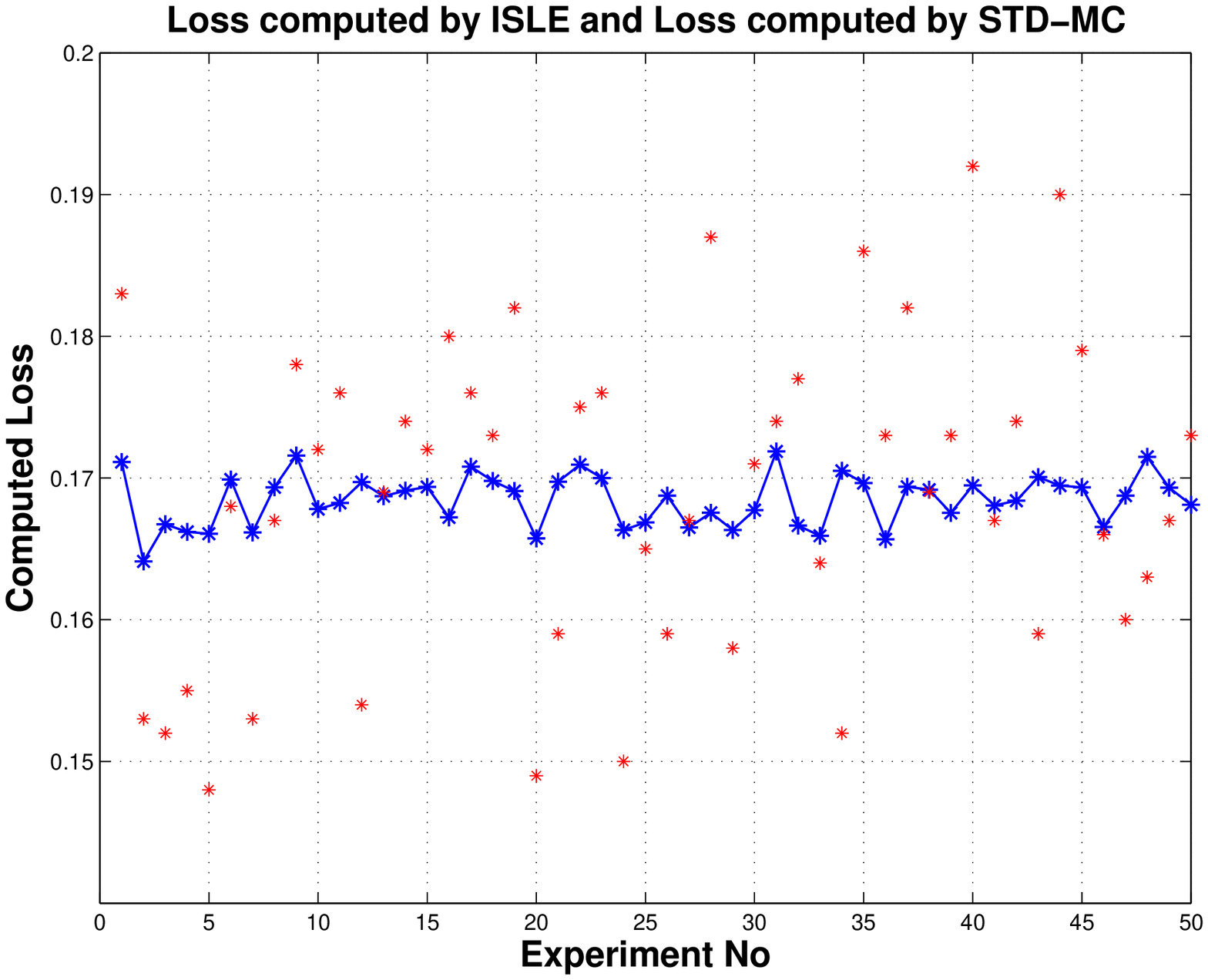}} \caption{Timing loss for {\sf Exp(\randCh, \varii)} (STD-MC and \slevv)}
\label{fig:plot1}
\end{figure}

The graph in \figg{plot1} shows the loss estimates obtained with the
STD-MC estimator and ISLE for all of the 50 experiment runs.  The
value of the ISLE estimator in each case is a lot closer to the mean
than that of the STD-MC estimator.  The variance reduction obtained by
the ISLE estimator over STD-MC is thus apparent from this graph. We
should note that for every loss estimate shown in \figg{plot1}, 1000
transistor-level simulations are performed for STD-MC, but the average
number of simulations for ISLE was only 213 over the 50 runs. Thus, if
a normalization (to be explained below) is done considering that the
errors for the two estimators are different, the {\gain} of ISLE over
STD-MC is found to be 179, theoretically given by \eq{gain1}. {\gain}
represents the ratio of the number of full circuit simulations
required by the two approaches to achieve the same error.

\begin{figure}
\centering \ForceWidth{2.5in}
\centerline{\BoxedEPSF{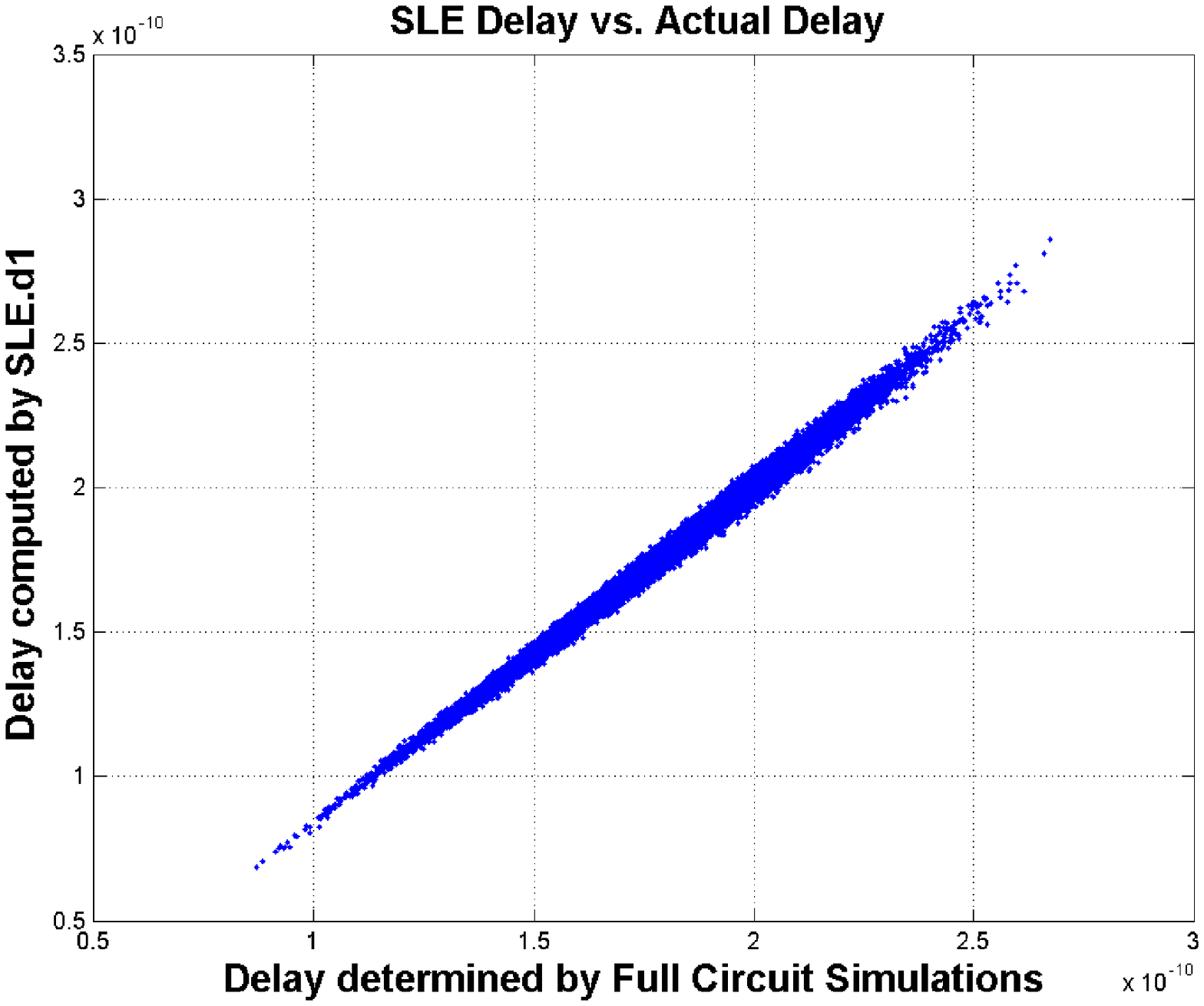}}
\caption{Scatter plot for {\sf Exp(\randCh, \varii)}, (\slev} \label{fig:plot2}
\end{figure}

\begin{figure}
\centering \ForceWidth{2.5in}
\centerline{\BoxedEPSF{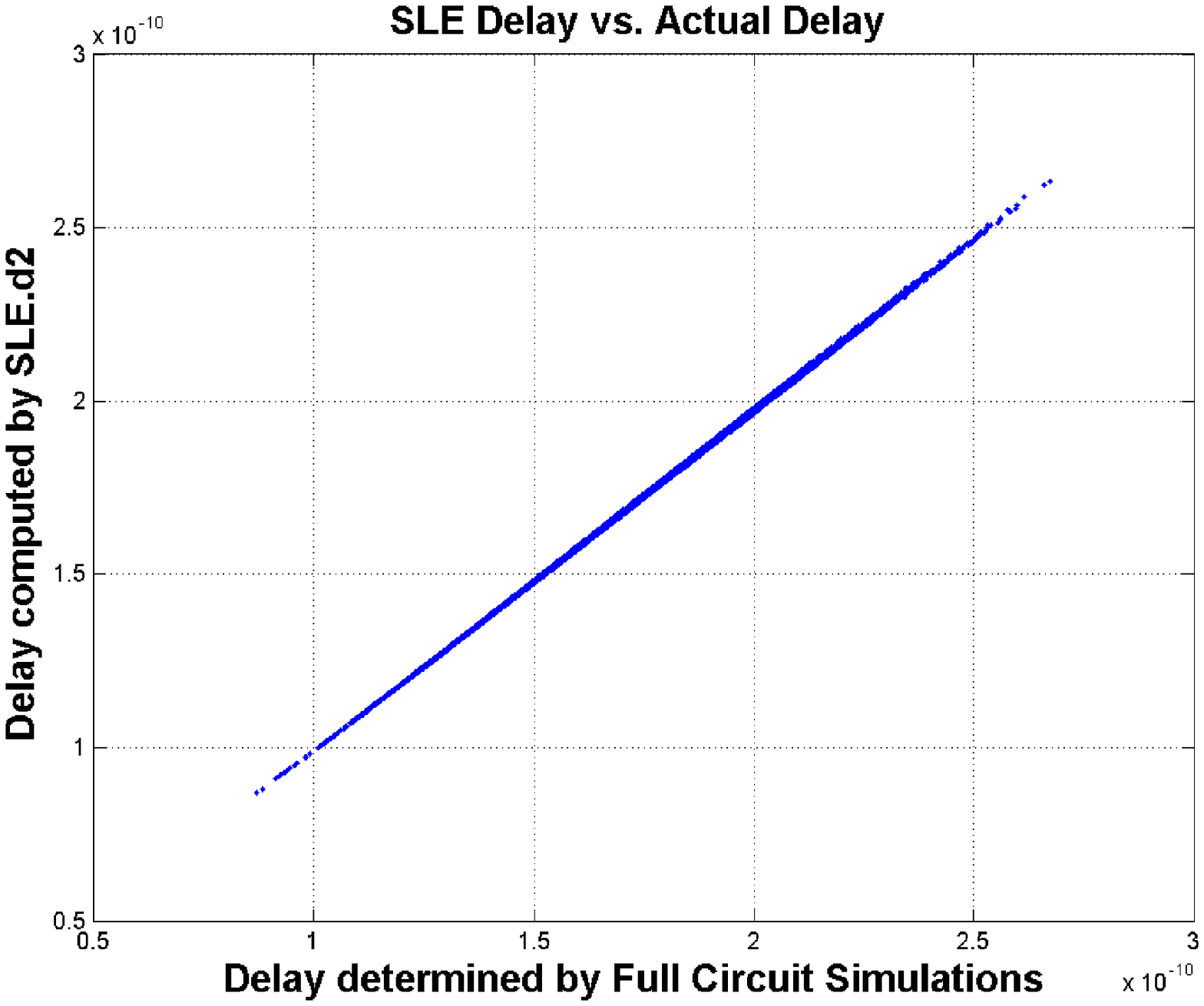}}
\caption{Scatter plot for {\sf Exp(\randCh, \varii)} (\slevv)} \label{fig:plot3}
\end{figure}

As discussed before, the accuracy of the SLE approximation is key in
order for it to facilitate IS for yield estimation. To gauge this
accuracy, the scatter plots in \figg{plot2} and \figg{plot3} show
circuit delays computed with the SLE formulas versus delay
computed with transistor-level circuit simulation. \figg{plot2} is for
$\slev$ and \figg{plot3} is for $\slevv$. As seen in these plots, both
versions of SLE formulas provide reliable delay estimates. However,
$\slevv$ is more accurate, and hence it results in a bigger {\gain} as
confirmed by the detailed results we present in the next section. This
comes at the cost of having statistical pre-characterizations for
parasitic delay $\paras$ and logical effort $\logeffort$ for all of
the gates in the library, in addition to the reference inverter delay
$\refinvdel$.

\begin{figure}
\centering \ForceWidth{2.5in}
\centerline{\BoxedEPSF{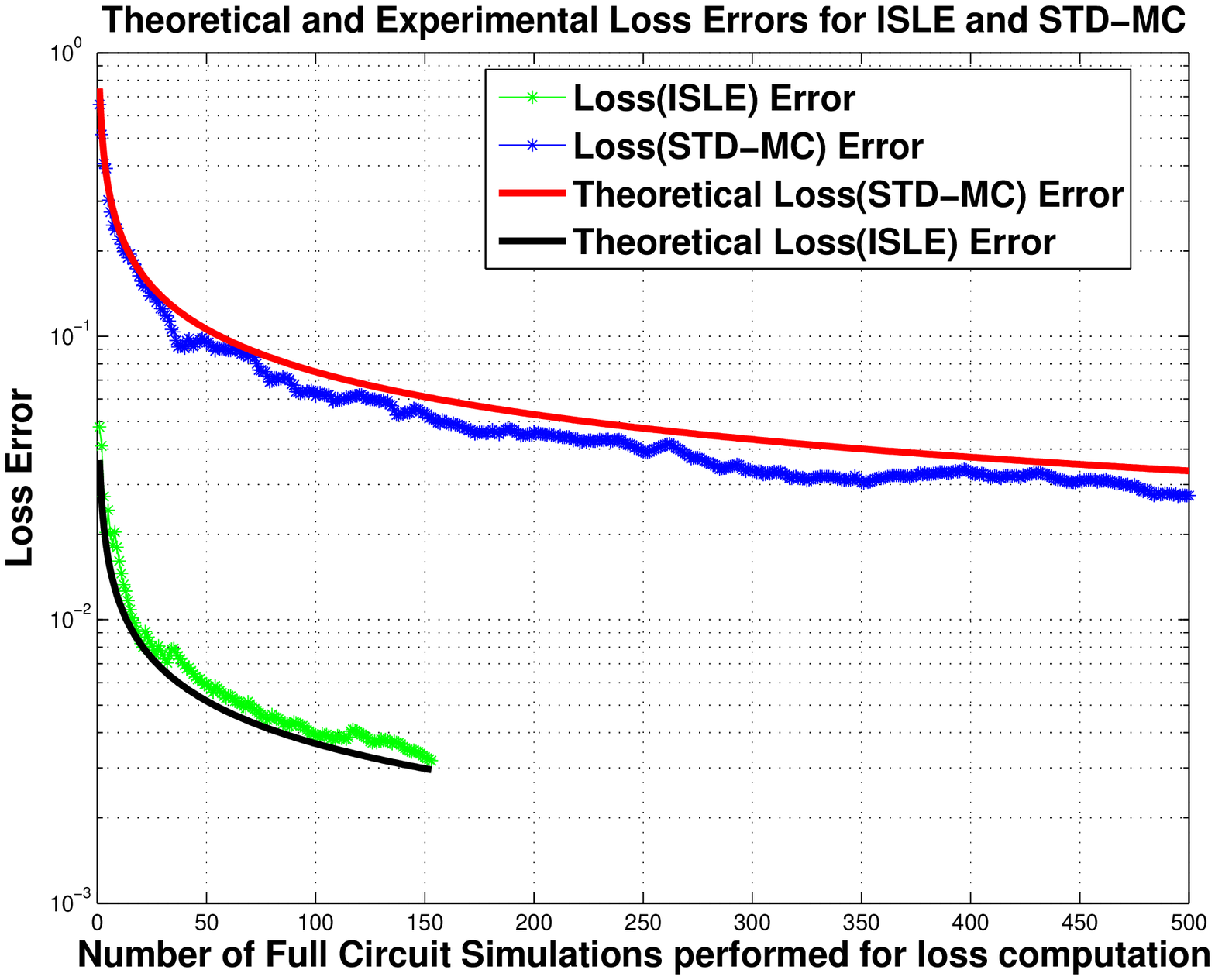}}
\caption{Log-Variance Plot for {\sf Exp(\randCh, \varii)} (\slevv)} \label{fig:plot4}
\end{figure}

\figg{plot4} confirms empirically that the $1/\sqrt{N}$ dependency of
error on the number of MC samples is as expressed by \eq{error-MC} and
\eq{error-IS-SLE} for the STD-MC and ISLE estimators, respectively.
The significant reduction in variance that ISLE provides
is also obvious in this graph. In this figure, a plot of loss error
versus the number of full circuit simulations is shown for both
estimators. The smooth curves in this plot were obtained using the
theoretical error formulas. The two other curves were computed using
data from the 50 runs, each of which explore sample set sizes ranging
from 1 to 500. As explained before, full circuit simulations are
performed at all of the sample points for the STD-MC estimator, but a
reduced number of simulations are needed for the ISLE estimator. We
observe the excellent match between the theoretical and experimental
error curves in this plot.


\begin{table*}[t]
\caption{Experimental Results Comparing Standard MC and ISLE}
\begin{tabular}[centering]{|l|c|c|c|c|c|c|c|c|c|c|c|}
\hline
           &   \multicolumn{ 3}{c|}{{\small \bf  Mean Loss}}      &           \multicolumn{ 3}{c|}{{\small \bf  Loss Error}}           &           \multicolumn{ 3}{c|}{{\small \bf Number of Ckt. Simulations}}           &            \multicolumn{ 2}{c|}{{\small \bf  Gain }}          \\
\hline
           &   \small \bf  SLE.d1 &   \small \bf  SLE.d2 &   \small \bf  STD-MC &   \small \bf  SLE.d1 &  \small  \bf  SLE.d2 &   \small \bf  STD-MC &   \small \bf  SLE.d1 &   \small \bf  SLE.d2 &   \small \bf  STD-MC &   \small \bf  SLE.d1 &   \small \bf  SLE.d2  \\
\hline
        {\rule[0mm]{0mm}{0mm}\small \bf \exper(\invCh, \vari)} &        \small  0.1394 &        \small  0.1394 &      \small    0.1395 &    \small      1.15e-3 &     \small     1.28e-3 &   \small       2.35e-2 &    \small      181 &       \small   181 &      \small    1000 &      \small \bf  2305 &       \small \bf 1866  \\
\hline
        {\rule[0mm]{0mm}{0mm}\small \bf  \exper(\invCh, \varii)} &        \small 0.1481 &        \small 0.1483 &      \small   0.1482 &     \small    8.15e-3 &      \small   1.72e-3 &    \small     1.87e-2 &     \small    211 &       \small  190 &       \small 1000 &     \small  \bf  25 &      \small  \bf 624 \\
\hline
        {\rule[0mm]{0mm}{0mm}\small \bf  \exper(\invCh, \variii)} &        \small 0.1535 &        \small 0.1537 &      \small   0.1538 &     \small    8.72e-3 &      \small   2.15e-3 &    \small     2.03e-2 &      \small   218 &        \small 196 &        \small 1000 &    \small  \bf   25 &     \small \bf   457  \\
\hline
        {\rule[0mm]{0mm}{0mm}\small \bf  \exper(\randCh, \vari)} &        \small 0.1575 &        \small 0.1575 &      \small   0.1576 &      \small   1.32e-3 &      \small   1.33e-3 &    \small     2.34e-2 &      \small   199 &        \small 199 &        \small 1000 &    \small   \bf  1589 &     \small  \bf  1549  \\
\hline
        {\rule[0mm]{0mm}{0mm}\small \bf  \exper(\randCh, \varii)}  &        \small 0.1686 &        \small 0.1684 &      \small   0.1688 &      \small   1.08e-2 &      \small   2.81e-3 &     \small    2.24e-2 &      \small   248 &        \small 211 &        \small 1000 &    \small  \bf   17 &      \small  \bf 300 \\
\hline
        {\rule[0mm]{0mm}{0mm}\small \bf  \exper(\randCh, \variii)}  &        \small 0.1739 &        \small 0.1741 &      \small   0.1743 &       \small  1.28e-2 &       \small  2.80e-3 &     \small    2.29e-2 &      \small   259 &        \small 217 &        \small 1000 &     \small  \bf  12 &      \small \bf  307 \\
\hline
\end{tabular}
\label{tab:tab1}
\vspace*{-0.3cm}
\end{table*}

\subsection{Results}
\label{sec:results}
\noindent
\tabb{tab1} present results obtained from six experiments with the
STD-MC estimator and our ISLE estimator based on both {\slev} and
{\slevv}. The mean loss and loss error values are computed over 50
independent runs as explained before. The {\gain} value reported in
the table represents the ratio of the number of full circuit
simulations required by the STD-MC and SLE estimators to achieve the
same error, theoretically given by \eq{gain1}.  {\sf Gain} here is
computed using the experimental data for the loss errors and the
actual number of circuit simulations reported in the table as follows:
\begin{equation}
\gain = \frac{N_{MC}}{N_{ISLE}}\:\frac{\errMC^2}{\errISLE^2}
\end{equation}
The second factor in the above formula is needed to perform a
normalization required in order to make a correction for the
difference in the errors achieved by the two estimators with the given
number of simulations. The ratio of the squares of errors is used
because of the $1/\sqrt{N}$ dependency of error.


The results in \tabb{tab1} show that both versions of our ISLE yield
estimator achieve significant cost reduction over the standard MC
estimator for the same error, in the range from one to four orders
of magnitude. As expected, ISLE based on {\slevv} performs better,
achieving two orders of magnitude cost reduction in the worst-case,
whereas the cost reduction achieved by ISLE based on {\slev} goes down
to 12 in the worst-case. The {\gain} data presented in \tabb{tab1}
shows that ISLE performs better for the {\invCh} circuit containing
only inverters.  This is due to the fact that the SLE delay formulas
are more accurate for inverters because of their use in the SLE
formalism as a delay reference. It should also be noted that the
performance of ISLE improves considerably when only one statistical
parameter is used, achieving a speed-up reaching three orders of
magnitude. When more (two or three) statistical parameters are
considered simultaneously, the performance degrades in comparison but
still above a respectable two orders of magnitude speed-up with ISLE
based on {\slevv}.


%% file: conc.tex
\noindent
We have demonstrated in this paper that Monte Carlo simulation in
conjunction with a novel variance reduction technique, {\em Importance
  Sampling with Stochastic Logic Effort}, can serve as an accurate yet
computationally viable timing yield estimation method. Numerous other
techniques for reducing the variance of Monte Carlo estimators have
been proposed in the Monte Carlo simulation
literature~\cite{KalosWhitlock86,Sokal97}. The stochastic logical
effort formalism can be used to facilitate other techniques for
variance reduction in Monte Carlo estimation, e.g., stratified
sampling, control variates, multicanonical Monte Carlo method, etc.
\cite{StratMC}, which we intend to explore in future work.


%% file: slemc.bbl
\begin{thebibliography}{1}

\bibitem{SchefferTAU2004}
L.~Scheffer.
\newblock The count of {M}onte {C}arlo.
\newblock In {\em ACM/IEEE International Workshop on Timing Issues in the
  Specification and Synthesis of Digital Systems (TAU)}, February 2004.

\bibitem{logeffortbook}
I.~Sutherland, B.~Sproull, and D.~Harris.
\newblock {\em Logical Effort: Designing Fast {CMOS} Circuits}.
\newblock Morgan Kaufmann, 1999.

\bibitem{sleTAU2006}
A.~Demir and S.~Tasiran.
\newblock Statistical logical effort: Designing for timing yield on the back of
  an envelope.
\newblock In {\em ACM/IEEE International Workshop on Timing Issues in the
  Specification and Synthesis of Digital Systems (TAU)}, February 2006.

\bibitem{KalosWhitlock86}
Malvin~H. Kalos and Paula~A. Whitlock.
\newblock {\em {Monte Carlo Methods, Volume 1, Basics}}.
\newblock Wiley, 1986.

\bibitem{StratMC}
P.~Glasserman, P.~Heidelberger, and P.~Shahabuddin.
\newblock Importance sampling and stratification for value-at-risk.
\newblock In {\em Proc. 6th Intl. Conference on Computational Finance}, pages
  7--24. MIT Press, May 28-31 1999.

\bibitem{vlsitech}
http://www.vlsitechnology.org/.
\newblock Cell Library, Release 8.1.

\bibitem{RabaeyELECTRON2003}
Y.~Cao, H.~Qin, R.~Wang, P.~Friedberg, A.~Vladimirescu, and J.~Rabaey.
\newblock Yield optimization with energy-delay constraints in low-power digital
  circuits.
\newblock In {\em IEEE Conference on Electron Devices and Solid-State
  Circuits}, December 2003.

\bibitem{Sokal97}
A.~D. Sokal.
\newblock Monte carlo methods in statistical mechanics: Foundations and new
  algorithms.
\newblock In P.~Cartier C.~DeWitt-Morette and A.~Folacci, editors, {\em
  Functional Integration: Basics and Applications (1996 Carg\`ese summer
  school)}. Plenum, 1997.

\end{thebibliography}
